\documentclass{epl}
\def\<{\langle }
\def\>{\rangle }
\def\kk{\>\!\>}
\def\bb{\<\!\<}
\def\Tr{\hbox{Tr} }
\def\map#1{\mathscr{#1}}
\def\sH{\mathcal H}
\usepackage{amsmath,mathrsfs}
\usepackage{amssymb}

\title{Superbroadcasting of conjugate quantum variables}
\shorttitle{Superbroadcasting of conjugate ...}
\author{Giacomo M. D'Ariano \inst{1} \and Paolo Perinotti \inst{1} 
\and Massimiliano F. Sacchi \inst{1,2}}
\shortauthor{G. M. D'Ariano, P. Perinotti,  M. F. Sacchi}

\institute{                    
  \inst{1} Dipartimento di Fisica ``A. Volta'' and CNISM, via Bassi 6,
I-27100 Pavia, Italy \\
  \inst{2} CNR - Istituto Nazionale per la Fisica della Materia, 
Unit\`a di Pavia, Italy.
}

\begin{document}

\maketitle

\begin{abstract}
  We consider the problem of broadcasting arbitrary states of
  radiation modes from $N$ to $M >N$ copies by a map that preserves
  the average value of the field and optimally reduces the total noise
  in conjugate variables. For $N \geq 2$ the broadcasting can be
  achieved perfectly, and for sufficiently noisy input states one can
  even purify the state while broadcasting---the so-called {\em
    superbroadcasting}. For purification (i.e.  $M\leq N$), the
  reduction of noise is independent of $M$.  Similar results are
  proved for broadcasting with phase-conjugation. All the optimal maps
  can be implemented by linear optics and linear amplification.
\end{abstract}

The impossibility of exact quantum cloning, namely copying the unknown
state of a quantum system to a larger number of copies \cite{noclon},
has stimulated the search for quantum devices that can emulate cloning
with the highest possible fidelity. After the simplest case of qubits
\cite{buzhill,gismass,bruss} many optimal cloners have been found, for
general finite-dimensional systems \cite{werner}, restricted sets of
input states \cite{darlop,darmacc}, and infinite-dimensional systems
such as harmonic oscillators---the so called continuous variables
cloners \cite{cerf}.  However, when considering mixed states, a less
stringent type of cloning transformation can be used---so-called {\em
  broadcasting}---in which the output copies are in a globally
correlated state whose local reduced states are identical to the input
states. This issue has been considered in Ref.  \cite{nobro}, where it
has been shown that broadcasting a single copy from a noncommuting set
of density matrices is always impossible.  Later, this result has been
considered in the literature as the generalization of the no-cloning
theorem to mixed states. However, more recently, for qubits an effect
called {\em superbroadcasting} \cite{prl} has been discovered, which
consists in the possibility of broadcasting the state while even
increasing the purity of the local state, for at least $N\ge 4$ input
copies, and for sufficiently short input Bloch vector (and even for
$N=3$ input copies for phase-covariant instead of universal covariant
broadcasting \cite{pra}).

In the present Letter, we study the broadcasting of conjugate quantum
variables $x_a=\frac {a +a^\dag }2$ and $y_a=\frac{a-a^\dag }{2i}$,
where $a$ and $a^\dag $ are the customary annihilation and creation
operators for the harmonic oscillator (or single-mode radiation
field), i.e. $[a,a^\dag ]=1$.  We look for the map that from $N$
uncorrelated states with the same complex amplitude 
provides $M>N$ states, while preserving the amplitude and optimally
reducing the total noise in conjugate quadratures.  We derive a bound
from the quantum limits on noise in {\em linear} amplifiers that can
be easily achieved experimentally.  We will show indeed that such a
bound cannot be overcome even for general nonlinear transformations
(i.e. allowing for arbitrary quantum operations). Explicit examples
will be given for displaced thermal states, which are equivalent to
coherent states that have suffered Gaussian noise.

As we will see, superbroadcasting is possible for continuous variables
for $N\geq 2$, namely one can produce a larger number $M$ of purified
copies at the output, locally on each use, and with the same amplitude
of the input copies $N$.  For displaced thermal states, e.g., $N$ to
$M$ superbroadcasting can be achieved for input thermal photon number
$\overline{n}_{in}> \frac{M-N}{M(N-1)}$.  For purification (i.e.
$M\le N$), quite surprisingly the purification rate is
$\overline{n}_{out}/\overline{n}_{in}=N^{-1}$, independently of $M$.
We mention that the particular case of $2\to1$ purification for noisy
coherent states has been reported in Ref. \cite{ula}. We will also
consider the optimal broadcasting with phase-conjugate output, showing
analogous effects.

From $N$ uncorrelated modes $a_0,a_1,...,a_{N-1}$ with
\begin{eqnarray}
\< a_ i \>=\alpha \;, \qquad 
\Delta x_{a_i} ^2 + \Delta y_{a_i} ^2 = \gamma _i \;, 
\label{ag}
\end{eqnarray}
a broadcasting transformation provides 
$M>N$ (generally correlated) 
modes $b_0,b_1,...,b_{M-1}$, with the same complex amplitude and 
noise $\Gamma $, i.e.  
\begin{eqnarray}
\< b_ i \>=\alpha \;, \qquad 
\Delta x_{b_i} ^2 + \Delta y_{b_i} ^2 = \Gamma 
\;,\label{bg}
\end{eqnarray} 
and we are looking for the minimal $\Gamma $. This can be obtained by
applying a fundamental theorem for linear
amplifiers: the sum of the uncertainties of conjugate quadratures of
an amplified mode with (power) gain $G$ is bounded as
follows \cite{caves}
\begin{eqnarray}
\Delta X_B ^2 +\Delta Y_B ^2  \geq G
(\Delta X_A ^2 +\Delta Y_A ^2  ) + \frac {|G \mp 1|}{2}
\;,\label{bou}
\end{eqnarray}
where the upper (lower) sign holds for phase-preserving
(phase-conjugating) amplifiers, and $A$ and $B$ denote the input and
the amplified mode, respectively. In fact, our transformation can be
seen as a phase-preserving amplification from mode $A=\frac {1}{\sqrt
  N}\sum _{i=0}^{N-1} a_i$ to mode $B=\frac {1}{\sqrt M}\sum
_{i=0}^{M-1} b_i$ with gain $G=\frac MN$, and hence Eq.  (\ref{bou})
should hold. Notice that generally for any mode $c$ one has 
\begin{eqnarray}
\Delta x_c ^2 +\Delta y_c ^2  = \frac 12 + \< c^\dag c \> -|\<c\>|^2
\;.\label{hol}
\end{eqnarray}
In the present case, since modes $a_i$ are uncorrelated, from
Eqs. (\ref{ag}) and (\ref{hol})  we have 
$\<A^\dag
A\> = \gamma +N |\alpha |^2 -\frac 12 $, 
with $\gamma =\frac 1N \sum _{i=0}^{N-1}\gamma _i$, 
whereas from Eqs. (\ref{bg}) and (\ref{hol})
\begin{eqnarray}
\<B^\dag B\> 
= \frac 1M \sum _{i,j=0}^{M-1} \<b^\dag _i b_j\>\leq 
\frac 1M \sum _{i,j=0}^{M-1} \sqrt{\<b^\dag _i b_i\> \<b^\dag _j b_j\> 
}
= M\left (\Gamma + |\alpha |^2 -\frac 12\right )
\;.
\label{ass2}
\end{eqnarray}
From Eqs. (\ref{bou}-\ref{hol}) we obtain the following bound for the 
noise $\Gamma$  
\begin{eqnarray}
\Gamma -\frac 12 \geq  \frac 1N \left 
(\gamma -\frac 12 \right ) +\frac {1}{N}-
\frac {1}{M}
\;.\label{gam}
\end{eqnarray}
Notice that $\gamma ,\Gamma \geq \frac 12$, due to the Heisenberg
uncertainty relations.  A similar derivation gives a bound for
purification, where $N > M$.  In such a case $G<1$, and one obtains
\begin{eqnarray}
\Gamma -\frac 12 \geq  \frac 1N \left (\gamma -\frac 12\right ) 
\;.\label{pur}
\end{eqnarray}
We will see that the $M$ output purified copies are, however,
correlated.
\par From the bound on phase-conjugating amplifiers (\ref{bou}), similarly  
it follows 
\begin{eqnarray}
\Gamma -\frac 12 \geq \frac{1}{N}\left (\gamma +\frac 12 \right )\;,\label{phc}
\end{eqnarray}
for phase-conjugating broadcasting (and purification).  The bound
(\ref{phc}) is independent of the number of output copies, and
corresponds to (\ref{gam}) for broadcasting in the limit $M\to\infty$.

The bound (\ref{gam}) for broadcasting can be achieved by the
following experimental setup. By means of a $N$-splitter the signal is
concentrated in one mode, whereas the other $N-1$ modes are discarded.
The mode is then amplified by a phase-insensitive amplifier with power
gain $G=\frac MN$. Finally, the amplified mode is mixed in a
$M$-splitter with $M-1$ vacuum modes.  In the concentration stage the
$N$ modes with amplitude $\<a \>=\alpha $ and noise $\Delta x^2
+\Delta y^2 =\gamma _i$ are reduced to a single mode with amplitude
$\sqrt N \alpha $ and noise $\gamma $. The amplification stage gives a
mode with amplitude $\sqrt M \alpha $ and noise $\gamma ' =\gamma
\frac MN + \frac {M}{2N} -\frac 12$.  Finally, the distribution stage
gives $M$ modes, with amplitude $\alpha $ and noise $\Gamma = \frac 1M
\left (\gamma ' +\frac {M-1}{2}\right )$ each. We notice that the
linear amplifier can be replaced by a beam-splitter, heterodyne
detection and feed-forward, as in Ref. \cite{ula2}.

\begin{figure}
\onefigure[width=5.5cm]{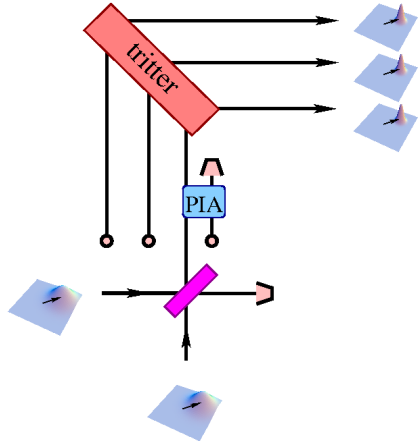}
\caption{Experimental scheme for optimal superbroadcasting 
  from 2 to 3 copies. The schemes makes use of a beam splitter, a
  phase-insensitive amplifier and a tritter (i.e.  two suitably
  balanced beam splitters). The output copies carry the same signal as
  the input, and are locally less noisy, the noise being confined
  into the correlations between them.}
\label{sch}
\end{figure}

The condition for superbroadcasting is given by $\Gamma < \gamma $,
namely $\gamma -\frac 1 2 \geq\frac{M-N}{M(N-1)}$, which can be true
for any $N>1$, and up to $M\le\infty$. Consider, for example, the case
of $N$ displaced thermal states $D(\alpha )\rho _ {\bar n}D^\dag
(\alpha )$ where
\begin{equation}
\rho_{\bar n} 
=\frac1{\bar n+1}\left(\frac{\bar n}{\bar n+1}\right)^{a^\dag
  a}\,, 
\label{therm}
\end{equation}
and $\bar n$ denotes the thermal photon number.  The output
state is given by $D(\alpha )^{\otimes M} \lambda D^\dag (\alpha
)^{\otimes M}$, with 
\begin{equation}
\lambda = 
\int\frac{M d^2\gamma}{\pi\bar n'} 
|\gamma\>\<\gamma|^{\otimes M} e^{-\frac{M|\gamma|^2}{\bar n'}}
\;,
\label{finalnumb}
\end{equation}
where $\bar n'=\frac{M(\bar n+1)}{N}-1$. Such a 
state is permutation-invariant and separable, with displaced thermal
state at each use, with thermal photon number 
\begin{equation}
\bar n''=\frac{\bar n'}{M}=\frac{\bar n}{N}+\frac{M-N}{MN}\,.
\label{super}
\end{equation}
The superbroadcasting condition (output purity higher than the input
one), is equivalent to require smaller thermal photon number at the
output than at the input, namely $\bar n\geq\frac{M-N}{M(N-1)}$. In
fact, $\gamma =\bar n +\frac 12$ and $\Gamma =\bar n'' +\frac 12$.
Notice that for $\bar n=0$ one has $N$ coherent states at the input,
and $\bar n''=\frac{M-N}{MN}$, namely one finds the optimal cloning
for coherent states of Ref. \cite{cerfbrauns}.  In Fig. \ref{sch} we
sketch the scheme for optimal $2$ to $3$ broadcasting. The
superbroadcasting effect arises for $\bar n > \frac 13$.

For achieving the optimal purification for any $M \leq N$, one simply
uses an $N$-splitter which concentrates the signal in one mode and
discards the other $N-1$ modes. Then by $N$-splitting with $N-1$
vacuum modes, one obtains $N$ purified signals (although correlated),
with equality in Eq. (\ref{pur}).
 
The bound for phase-conjugating broadcasting and purification
(\ref{phc}) can be obtained by $N$-splitting, heterodyne
measurement on one of the modes, and preparation of $M$ coherent
states with amplitude $\alpha = \frac {\alpha ^* _o} {\sqrt N} $,
where $\alpha _o$ denotes the outcome of the measurement.

We would like to stress that the bounds (\ref{gam}), (\ref{pur}), and
(\ref{phc}) hold for any state and relies on the theorem of the added
noise in {\em linear} amplifiers, namely only linear transformations
of modes are considered.  Hence, in principle, these bounds might be
violated when considering a restricted set of states and allowing for
more exotic and nonlinear transformations.

In the following we give a rigorous proof that these bounds indeed
cannot be overcome by any quantum transformation. Let us consider a
generic state $\Xi _\alpha $ of $N$ uncorrelated modes with noise
$\gamma _i$, and $\langle a_i \rangle =\alpha $ for all modes. Then,
$\Xi _\alpha $ can be written as $D(\alpha)^{\otimes N}\Xi _0 D^\dag
(\alpha)^{\otimes N}$, where $D(\alpha)=\exp(\alpha a^\dag-\alpha^*
a)$ denotes the displacement operator and $\Xi_0=\otimes
_{i=0}^{N-1}\xi _i$ is the tensor product of $N$ states, each with
zero amplitude (i.e., for a single-mode radiation field, zero average
value of the field) and noise $\gamma _i$. We look for a broadcasting
map $\map B$ that preserves the unknown amplitude on each copy
\begin{eqnarray}
\Tr [b_i \,\map B (D(\alpha)^{\otimes N}\Xi _0
D^\dag (\alpha)^{\otimes N} ]
= \alpha \;,\label{ss}
\end{eqnarray}
for all $i \in [0,M-1]$ and complex $\alpha $,  such that each copy 
has minimal noise $\Gamma $, where, using  Eq. (\ref{hol}) 
\begin{eqnarray}
\Gamma 
=\frac 12 + \Tr[b^\dag _i b_i \,
\map B (D(\alpha)^{\otimes N}\Xi _0 
D^\dag (\alpha)^{\otimes N}) ] - |\alpha |^2\;.\label{nn}
\end{eqnarray}
The optimal broadcasting map can be searched among {\em covariant} 
 maps $\map B$ that satisfy for all $\sigma $ and $\alpha $ \cite{nota}
\begin{equation}
\map B(D(\alpha)^{\otimes N}\sigma D^\dag (\alpha)^{\otimes
  N})=D(\alpha)^{\otimes M}\map B(\sigma )
D^\dag (\alpha)^{\otimes M}\,.
\nonumber \label{cov}
\end{equation}
It is useful to consider the Choi-Jamio\l kowski bijective
correspondence of completely positive (CP) maps $\map B$ from
$\sH_\mathrm{in}$ to $\sH_\mathrm{out}$ and positive operators $
R_{\map B}$ acting on $\sH_\mathrm{out}\otimes\sH_\mathrm{in}$, which
is given by the following relations
\begin{equation}
\begin{split}
  &R_{\map B}=\map B\otimes \map I(|\Omega\>\<\Omega|)\;,\\
  &\map B(\rho)=\Tr_\mathrm{in}[(I_\mathrm{out}\otimes\rho^\tau )R_{\map
    B}]\;,
\end{split}
\end{equation}
where $|\Omega\>=\sum_{n=0}^{\infty}|\psi_n\>|\psi_n\>$ is a maximally
entangled vector of $\sH_\mathrm{in} ^{\otimes 2}$, and $X^\tau $ denotes
transposition of $X$ in the basis $|\psi_n\>$. In terms of the
operator
$R_{\map B}$ the covariance property 
can be written as
\begin{eqnarray}
[R_{\map B},D(\alpha )^{\otimes M}\otimes 
D(\alpha ^*)^{\otimes N}]=0\;,\qquad
\forall \alpha \in {\mathbb C}\;,\label{com}
\end{eqnarray}
and conditions (\ref{ss}) and (\ref{nn}) are equivalent to 
\begin{eqnarray}
&&
\Tr [b_i \otimes \Xi _0 ^\tau  R_{\map B}]= 0\;,\label{ss3}
\\& & 
\Gamma =\frac 12 + \Tr[b^\dag _i b_i \,\otimes \Xi _0 ^\tau   
R_{\map B}]\;.\label{ss33}
\end{eqnarray}
In order to deal with the covariance constraint we introduce the
multisplitter operators $U_a$ and $U_b$, that satisfy
\begin{eqnarray}
U_a a_k U^\dag _a =\frac 1{\sqrt N}
\sum_{l=0}^{N-1}e^{\frac{2\pi i kl}N} a_{l} \;,\qquad   
U_b b_k U^\dag _b =\frac 1{\sqrt M} \sum_{l=0}^{M-1}
e^{\frac{2\pi i
    kl}M}b_l\;, 
\end{eqnarray}
and the  squeezing transformation
$S_{a_0b_0}$ defined by 
\begin{equation}
 S_{a_0 b_0}a_0^\dag S_{a_0b_0}^\dag =\mu a_0^\dag -\nu b_0 \;,
\qquad 
  S_{a_0b_0}b_0S_{a_0b_0}^\dag =\mu b_0 -\nu a_0^\dag\;,
\label{ampli}
\end{equation}
with $\mu = \sqrt{\frac {M}{(M-N)}}$ and $\nu =\sqrt{\frac N{M-N}}$.  Condition
(\ref{com}) then becomes
\begin{eqnarray}
[S^\dag _{a_0b_0}(U^\dag_b\otimes U^\dag _a) R_{\map B} (U_b\otimes
U_a) S_{a_0b_0},
D_{b_0}(\alpha )]=0\;.\label{commrel}
\end{eqnarray}
Hence, upon introducing an operator $B$ of modes
$b_1,...,b_{M-1},a_0,...,a_{N-1}$, the operator $R_{\map B}$ can be
written in the form
\begin{eqnarray}
R_{\map B}= (U_b\otimes U_a) S_{a_0b_0} (I_{b_0}\otimes B)
S^\dag _{a_0b_0}(U^\dag_b\otimes U^\dag _a).
\end{eqnarray}
Notice that $R_{\map B}\geq 0$ is equivalent to $B\geq 0$. The further
condition that $\map B$ is trace-preserving in terms of $R_{\map B}$
becomes $\Tr_{b}[R_{\map B}]=I_{a}$, where $b$ and $a$ denotes
collectively all output and input modes.
This condition is verified iff
\begin{eqnarray}
\hbox{Tr}_{b \backslash b_0,a_0}[B]=\nu ^2
I_{a\backslash a_0}\;,
\label{trb}
\end{eqnarray}
where $a\backslash a_i$ denote all the input modes except $a_i$, and
similarly for $b\backslash b_i$.

Consider now the expectation value of the total number of photons of
the $M$ clones $W=\Tr[\sum _{l=0}^{M-1} b_l^\dag b_l \map B(\Xi _0)
]$.  Since the multisplitter preserves the total number of photons we
have
\begin{eqnarray}
W&=& 
\Tr\left [\left(\sum _{l=0}^{M-1} b_l^\dag b_l 
\otimes  U_a^\dag \Xi _0 ^\tau   U_a 
\right) S_{a_0b_0}(I_{b_0}\otimes B) S^\dag _{a_0 b_0}\right]
\nonumber \\&  
\geq  & \Tr\left [\left(b_0^\dag b_0 
\otimes  U_a^\dag \Xi  _0 ^\tau   U_a 
\right) S_{a_0b_0}(I_{b_0}\otimes B) S^\dag _{a_0 b_0}\right]
\;. 
\label{vutot}
\end{eqnarray}
Using the relation which hold for any state $\sigma $
\begin{eqnarray}
\Tr_{b_0} [ S^\dag _{a_0 b_0} (b^\dag _0 b_0
\otimes \sigma ) S_{a_0b_0} ] =  \frac{1}{\nu ^4}
(a_0^\dag a_0 \otimes \Tr _{a_0 } [\sigma  ] 
+ \ \mu ^2 I_{a_0}\otimes 
\Tr _{a_0}[a_0^\dag a_0 \sigma ] +I_{a_0}\otimes \Tr _{a_0}[\sigma
])\;,
\label{squiquiz}
\end{eqnarray}
along with condition (\ref{trb}),  
continuing from Eq. (\ref{vutot}) we obtain 
\begin{eqnarray}
W&=& \frac{\mu ^2 \Tr [a_0 ^\dag a_0 \, U_a^\dag \Xi _0^\tau   U_a 
] +1}
{\nu ^2} =
 \frac{\mu ^2 \frac 1 N 
\Tr [\sum _{i,j=0}^{N-1}a_i ^\dag a_j  \, \Xi _0^\tau  ]
+1}{\nu ^2} \nonumber \\& =& 
\frac{M}{N}\left (\gamma -\frac 12 \right ) +\frac{M-N}{N}
\;,\label{gam2}
\end{eqnarray}
which, from Eq. (\ref{ss33}), allows one to recover the bound
(\ref{gam}). With the choice $B=\nu ^2 \, |0\>\<0|_{b\backslash
  b_0}\otimes|0\>\<0|_{a_0}\otimes I_{a\backslash a_0}$
one can check that both the bound in Eq. (\ref{gam2}) is
achieved and Eq. (\ref{ss3}) is satisfied. In fact, such a choice of
$B$   gives a map that produces $M$ identical clones $D(\alpha )\rho
D^\dag (\alpha )$ with 
\begin{eqnarray}
\rho  = \int \frac{d^2 \alpha }{\pi}\,e ^{-\frac{|\alpha |^2}{2}(\frac
  1N -\frac 2M +1)}\, 
\{\Tr [\Xi_0  D^\dag  (\alpha  / N)^{\otimes N}]  \, D(\alpha )
\;.
\nonumber \label{rhogen}
\end{eqnarray}
The proof of the bound (\ref{pur}) for purification (for $M<N$
\cite{nota2}) can be obtained by analogous derivation, where now
$\mu=\sqrt{\frac{N}{N-M}}$ and $\nu=\sqrt{\frac{M}{N-M}}$, while the
trace-preserving condition becomes $\Tr_{b}[B]=\mu ^2 I_{a\backslash a_0}$. The
map corresponding to $B=\mu ^2 \,|0\>\<0|_{b}\otimes I_{a\backslash a_0}$ 
achieves the bound (\ref{pur}), and produces $M$ purified copies
$D(\alpha )\rho D^\dag (\alpha )$ with 
\begin{eqnarray}
\rho  = \int \frac{d^2 \alpha }{\pi}\,e ^{-\frac{|\alpha |^2}{2}(1- \frac
  1N)}\, 
\{\Tr [\Xi _0 D^\dag  (\alpha  / N)] \}^N \, D(\alpha )
\;.\nonumber 
\end{eqnarray}

A covariant phase-conjugating broadcasting map $\map C$ satisfies for
all $\sigma  $ and $\alpha $
\begin{equation}
\map C (D(\alpha)^{\otimes N}\sigma D^\dag (\alpha)^{\otimes N})=
D^*(\alpha)^{\otimes M}\map C(\sigma  )D^\tau (\alpha)^{\otimes M}
\nonumber
\end{equation}
which, in terms of $R_{\map C}$, corresponds to
$[D(\alpha)^{*\otimes(M+N)},R_{\map C}]=0$.  Introducing the
beam-splitter transformation
\begin{equation}
 U_{a_0b_0}b_0 U^\dag_{a_0b_0}=\eta b_0+\theta a_0 \,,\qquad 
U_{a_0b_0}a_0 U^\dag_{a_0b_0}=-\theta b_0+\eta a_0\;,
\end{equation}
with $\eta=\sqrt{\frac{M}{M+N}}$ and $\theta=\sqrt{\frac{N}{M+N}}$, 
the covariance relation gives $R_{\map C}$ of the form 
\begin{equation}
R_{\map C}= U_b \otimes
U_a  U_{a_0b_0} (I_{b_0}\otimes C)
U^\dag _{a_0b_0 }U^\dag_b \otimes U^\dag _a,
\end{equation}
where $C$ is an operator of modes $b_1,\dots,
b_{M-1},a_0,\dots,a_{N-1}$, with the trace-preserving condition $
\Tr_{b\backslash b_0,a_0}[C]=\theta^2I_{a\backslash a_0}$.  The proof
of the bound (\ref{phc}) is analogous to the case for the broadcasting
map, where one just replaces $U_{a_0b_0}$ with $S_{a_0b_0}$. The bound
can be achieved by $C=\theta^2 \, |0\>\<0|_{b\backslash
  b_0}\otimes|0\>\<0|_{a_0}\otimes I_{a\backslash a_0}$,
and the state of the clones are given by $D^*(\alpha )\rho D^\tau
(\alpha )$, with 
\begin{equation}
\rho  = \int \frac{d^2 \alpha }{\pi}\,e ^{-\frac{|\alpha |^2}{2}(1 +
  \frac  1N)}\, 
\{\Tr [\Xi _0 D^\dag  (\alpha  / N)] \}^N \, D(\alpha )\;.
\nonumber 
\end{equation}

In conclusion, we showed the optimal $N$ to $M$ phase-preserving/phase
conjugating broadcasting and purification maps for continuous
variables. For $N \geq 2$, the superbroadcasting can be achieved,
namely $M>N$ copies can be obtained along with a reduction of noise in
conjugate variables. Since the noise cannot be removed without
violating the quantum data processing theorem, the price to pay for
having higher purity at the output is that the copies are correlated.
Essentially noise is moved from local states to their correlations,
and the superbroadcasting channel that we presented does this
optimally. All the optimal maps can be easily implemented by linear
optics and linear amplification (or beam-splitting and feed-forward).
Superbroadcasting is relevant for foundations, opening new
perspectives in the understanding of correlations and their interplay
with noise, and may be also promising from a practical point of view,
for communication tasks in the presence of noise.

\acknowledgments 
This work has been supported by Ministero Italiano dell'Universit\`a e
della Ricerca (MIUR) through FIRB (bando 2001) and PRIN 2005.


\begin{thebibliography}{0}
\bibitem{nota} In fact, for any map $\map B$, one can construct a
  covariant one $\tilde {\map B}$ by averaging over the group, and
  still satisfying Eqs. (\ref{ss}) and (\ref{nn}).  Actually, being the
  group noncompact, a limit procedure has to be taken, e.g.
\begin{eqnarray}
\tilde {\map B }(\sigma )=\lim _{\Delta \rightarrow \infty }\int \frac {d^2
  \alpha }{\pi \Delta ^2}\,e^{-\frac {|\alpha  |^2}{\Delta ^2}} 
D^\dag
(\alpha ) ^{\otimes M}\map B [D(\alpha)^{\otimes N}\sigma   
D^\dag (\alpha)^{\otimes
  N}] D(\alpha)^{\otimes M}
\;\label{}
\end{eqnarray}
\bibitem{nota2} The proof for the case $M=N$ is technically slighty
  different, since the squeezing operator $S_{a_0b_0}$ is ill-defined
  in this case. By introducing the EPR states $|D(\beta)\kk =
  \sum_{n=0}^{\infty}(D(\beta) \otimes I )|n\>|n\>$, the covariance
  condition implies 
\begin{eqnarray}
  R_{\map B}=(U_b\otimes U_a)\int\frac{d^2\gamma}\pi|D(\gamma)\kk\bb
  D(\gamma)|_{a_0b_0} 
\otimes \Delta_{a\backslash a_0,b\backslash b_0}(\gamma)
(U_b^\dag\otimes U_a^\dag),
\end{eqnarray}
while the  trace-preserving constraint is given by 
$\int d^2\gamma \,  \Tr_{b\backslash b_0}[\Delta_{a\backslash a_0,b\backslash b_0}(\gamma)]=\pi
I_{a\backslash a_0}$. 
Then, the bound is proved as before, and is achieved for 
$\Delta_{a\backslash a_0,b\backslash b_0}(\gamma)=\pi\delta^2(\gamma)I_{a\backslash a_0}\otimes 
|0\>\<0|_{b\backslash b_0}$.
\end{thebibliography}
\end{document}